\let\useblackboard=\iftrue
%
%
\newfam\black
\input harvmac.tex
\noblackbox

\def\Title#1#2{\rightline{#1}
\ifx\answ\bigans\nopagenumbers\pageno0\vskip1in%
\baselineskip 15pt plus 1pt minus 1pt
\else
\def\listrefs{\footatend\vskip 1in\immediate\closeout\rfile\writestoppt
\baselineskip=14pt\centerline{{\bf References}}\bigskip{\frenchspacing%
\parindent=20pt\escapechar=` \input
refs.tmp\vfill\eject}\nonfrenchspacing}
\pageno1\vskip.8in\fi \centerline{\titlefont #2}\vskip .5in}

\ifx\answ\bigans\def\tcbreak#1{}\else\def\tcbreak#1{\cr&{#1}}\fi
\useblackboard
\message{If you do not have msbm (blackboard bold) fonts,}
\message{change the option at the top of the tex file.}
\font\blackboard=msbm10 scaled \magstep1
\font\blackboards=msbm7
\font\blackboardss=msbm5
\textfont\black=\blackboard
\scriptfont\black=\blackboards
\scriptscriptfont\black=\blackboardss

\else

\fi
\def\yboxit#1#2{\vbox{\hrule height #1 \hbox{\vrule width #1
\vbox{#2}\vrule width #1 }\hrule height #1 }}
\def\fillbox#1{\hbox to #1{\vbox to #1{\vfil}\hfil}}
\def\ybox{{\lower 1.3pt \yboxit{0.4pt}{\fillbox{8pt}}\hskip-0.2pt}}
\def\np#1#2#3{Nucl. Phys. {\bf B#1} (#2) #3}
\def\pl#1#2#3{Phys. Lett. {\bf #1B} (#2) #3}

\def\physrev#1#2#3{Phys. Rev. {\bf D#1} (#2) #3}

\def\comments#1{}

\def\half{{1\over 2}}

\def\CM{{\cal M}}

\def\II{\relax{I\kern-.07em I}}

\def\IZ{\relax\ifmmode\mathchoice
{\hbox{\cmss Z\kern-.4em Z}}{\hbox{\cmss Z\kern-.4em Z}}
{\lower.9pt\hbox{\cmsss Z\kern-.4em Z}}
{\lower1.2pt\hbox{\cmsss Z\kern-.4em Z}}\else{\cmss Z\kern-.4em
Z}\fi}
\def\IB{\relax{\rm I\kern-.18em B}}
\def\IC{{\relax\hbox{$\inbar\kern-.3em{\rm C}$}}}
\def\ID{\relax{\rm I\kern-.18em D}}
\def\IE{\relax{\rm I\kern-.18em E}}
\def\IF{\relax{\rm I\kern-.18em F}}
\def\IG{\relax\hbox{$\inbar\kern-.3em{\rm G}$}}
\def\IGa{\relax\hbox{${\rm I}\kern-.18em\Gamma$}}
\def\IH{\relax{\rm I\kern-.18em H}}
\def\II{\relax{\rm I\kern-.18em I}}
\def\IK{\relax{\rm I\kern-.18em K}}
\def\IP{\relax{\rm I\kern-.18em P}}

\font\cmss=cmss10 \font\cmsss=cmss10 at 7pt
\def\IR{\relax{\rm I\kern-.18em R}}

\def\IZ{\relax\ifmmode\mathchoice
{\hbox{\cmss Z\kern-.4em Z}}{\hbox{\cmss Z\kern-.4em Z}}
{\lower.9pt\hbox{\cmsss Z\kern-.4em Z}}
{\lower1.2pt\hbox{\cmsss Z\kern-.4em Z}}\else{\cmss Z\kern-.4em
Z}\fi}
\def\IB{\relax{\rm I\kern-.18em B}}
\def\IC{{\relax\hbox{$\inbar\kern-.3em{\rm C}$}}}
\def\ID{\relax{\rm I\kern-.18em D}}
\def\IE{\relax{\rm I\kern-.18em E}}
\def\IF{\relax{\rm I\kern-.18em F}}
\def\IG{\relax\hbox{$\inbar\kern-.3em{\rm G}$}}
\def\IGa{\relax\hbox{${\rm I}\kern-.18em\Gamma$}}
\def\IH{\relax{\rm I\kern-.18em H}}
\def\II{\relax{\rm I\kern-.18em I}}
\def\IK{\relax{\rm I\kern-.18em K}}
\def\IP{\relax{\rm I\kern-.18em P}}

\font\cmss=cmss10 \font\cmsss=cmss10 at 7pt
\def\IR{\relax{\rm I\kern-.18em R}}

\def\tilde{\widetilde}
\def\frac#1#2{{{#1} \over {#2}}}

\def\tN{\tilde{N}}

\Title{\vbox{\baselineskip12pt\hbox{hep-th/9709118}
\hbox{IASSNS-HEP-97/99, LBNL-40723, RU-97-72}
\hbox{SLAC-PUB-7648, UCB-PTH-97/45}
}}
{\vbox{\centerline{Matrix Description of $(1,0)$ Theories }
\centerline{}
\centerline{in Six Dimensions}}}

\centerline{Ofer Aharony$^1$, Micha Berkooz$^2$, Shamit Kachru$^3$, 
and Eva Silverstein$^4$}
\smallskip
\smallskip
\smallskip
\centerline{$^1$ Department of Physics and Astronomy}
\centerline{Rutgers University }
\centerline{Piscataway, NJ 08855-0849, USA}
\smallskip
\smallskip
\smallskip
\centerline{$^2$ School of Natural Sciences}
\centerline{Institute for Advanced Study}
\centerline{Princeton, NJ 08540, USA}  
\smallskip
\smallskip
\smallskip
\centerline{$^3$ Department of Physics}
\centerline{University of California at Berkeley}
\centerline{Berkeley, CA 94720, USA}
\smallskip
\smallskip
\smallskip
\centerline{$^4$ Stanford Linear Accelerator Center}
\centerline{Stanford University}
\centerline{Stanford, CA 94309, USA}
\bigskip
\bigskip
\noindent
We propose descriptions of interacting $(1,0)$ supersymmetric 
theories without gravity
in six dimensions in the infinite momentum frame.  
They are based on the large $N$ limit of
quantum mechanics or $1+1$ dimensional field theories with
$SO(N)$ gauge group and four supercharges.  
We argue that this formulation allows for a concrete description of
the chirality-changing phase transitions which connect $(1,0)$ theories with
different numbers of tensor multiplets.


\Date{September 1997}

\lref\berd{M. Berkooz and M. Douglas, ``Five-branes in M(atrix) Theory,''
Phys. Lett. {\bf B395} (1997) 196, hep-th/9610236.}%

\lref\blackholes{See, e.g., A. Strominger and C. Vafa,
``Microscopic Origin of the Bekenstein/Hawking Entropy,''
Phys. Lett. {\bf B379} (1996) 99, hep-th/9601029;
C. Callan and J. Maldacena, ``D-brane
Approach to Black Hole Quantum Mechanics,''
Nucl. Phys. {\bf B472} (1996) 591, hep-th/9602043.}

\lref\rohm{R. Rohm and E. Witten, \lq\lq Antisymmetric Tensor Field
in Superstring Theory,'' Ann. Phys. {\bf 170} (1986) 454.}

\lref\moorelitt{A. Losev, G. Moore, and 
S. Shatashvili, ``M$\&$m's,'' hep-th/9707250.}

\lref\finite{See e.g. P. Howe and G. Papadopoulos, \lq\lq Finiteness 
and Anomalies
in (4,0) Supersymmetric Sigma Models,'' hep-th/9203070, Nucl. Phys.
{\bf B381} (1992) 360, and references therein.}

\lref\douglas{M.R. Douglas, ``Gauge Fields and D-branes,'' hep-th/9604198.}%

\lref\hw{P. Ho\v rava and E. Witten, ``Heterotic and Type I Dynamics from
Eleven Dimensions,'' hep-th/9510209, \np{460}{1996}{506}.}%

\lref\abkss{O. Aharony, M. Berkooz, S. Kachru, N. Seiberg, and E. 
Silverstein,  
``Matrix Description of Interacting Theories in Six Dimensions,'' 
hep-th/9707079.}%

\lref\ndouglas{M.R. Douglas, ``Branes Within Branes,'' hep-th/9512077.}
\lref\bss{T. Banks, N. Seiberg and E. Silverstein, ``Zero and One
Dimensional Probes with $N=8$ Supersymmetry,'' hep-th/9703052,
\pl{401}{1997}{30}.}

\lref\bfss{T. Banks, W. Fischler, S. Shenker, and L. Susskind, 
``M theory as a Matrix Model:  A Conjecture,'' hep-th/9610043,
\physrev{55}{1997}{112}.}%

\lref\susskind{L. Susskind, ``Another Conjecture about M(atrix)
Theory,'' hep-th/9704080.}%

\lref\lambert{N.D. Lambert, ``D-brane Bound States and the Generalized
ADHM Construction,'' hep-th/9707156.}%

\lref\gstring{N. Seiberg, ``New Theories in Six Dimensions and
Matrix Description of M-theory on $T^5$ and $T^5/\IZ_2$,''
hep-th/9705221.}%
\lref\dvv{R. Dijkgraaf, E. Verlinde and H. Verlinde, ``BPS Spectrum of
the Fivebrane and Black Hole Entropy,'' hep-th/9603126,
\np{486}{1997}{77}; ``BPS Quantization of the Fivebrane,''
hep-th/9604055, \np{486}{1997}{89}.}%
\lref\dvvmarch{R. Dijkgraaf, E. Verlinde and H. Verlinde, ``5D Black
Holes and Matrix Strings,'' hep-th/9704018.}%

\lref\wittentwoz{E. Witten, ``Some Comments on String Dynamics,''
hep-th/9507121, Contributed to STRINGS 95: 
Future Perspectives in String Theory, Los Angeles, CA, 13-18 Mar 
1995.}

\lref\strom{A. Strominger, ``Open P-branes,'' hep-th/9512059,
\pl{383}{1996}{44}.}%

\lref\ginsparg{P. Ginsparg, ``Comment on Toroidal Compactification of
Heterotic Superstrings,'' \physrev{35}{1987}{648}.}%

\lref\dkps{M.R. Douglas, D. Kabat, P. Pouliot, and S. Shenker, 
``D-branes and Short Distances in String Theory,''
Nucl. Phys. {\bf B485} (1997) 85, hep-th/9608024.}%

\lref\dg{J. Distler and B. Greene, ``Aspects of $(2,0)$ String
Compactifications,'' \np{304}{1988}{1}.}%

\lref\newwitten{E. Witten, ``On the Conformal Field Theory of the Higgs
Branch,'' hep-th/9707093.}%

\lref\polwit{J. Polchinski and E. Witten, ``Evidence for Heterotic -
Type I String Duality,'' hep-th/9510169, \np{460}{1996}{525}.}%

\lref\vafwit{C. Vafa and E. Witten, ``A Strong Coupling Test of
S-Duality,'' Nucl. Phys. {\bf B431} (1994) 3, hep-th/9408074.}

\lref\orb{L. Dixon, J. Harvey, C. Vafa, and E. Witten,
``String on Orbifolds,'' Nucl. Phys. {\bf B261} (1985) 678.}

\lref\clifford{C.V. Johnson, work in progress.}

\lref\motls{L. Motl and L. Susskind, ``Finite $N$ Heterotic Matrix
Models and Discrete Light Cone Quantization,'' hep-th/9708083.}%

\lref\newken{K. Intriligator, ``New String Theories in Six Dimensions
via Branes at Orbifold Singularities,'' hep-th/9708117.}%

\lref\wittenfourone{E. Witten, ``Branes and the Dynamics of QCD,''
hep-th/9706109.}%

\lref\newlowe{D.A. Lowe, ``$E_8 \times E_8$ Small Instantons in Matrix
Theory,'' hep-th/9709015.}%

\newsec{Introduction}

In the past few years, large classes of interacting superconformal
field theories with between 4 and 16 supercharges have been discovered in
three, four, five and six space-time dimensions.
Most of these theories are not described by weakly-coupled Lagrangians,
and there is not even a known Lagrangian which flows to them
in many cases. Therefore,
we require a different approach to analyze them.  
This is an interesting abstract problem in itself, and it is rendered more
urgent by the many applications these theories have
in M theory.  In the matrix formulation
of M theory \bfss\ these theories are relevant for compactifications
on four dimensional spaces 
\nref\rozali{M. Rozali, ``Matrix Theory and U Duality in Seven
Dimensions,'' hep-th/9702136.}%
\nref\brs{M. Berkooz, M. Rozali and N. Seiberg,  ``Matrix Description
of M theory on $T^4$ and $T^5$,'' hep-th/9704089.}%
\nref\br{M. Berkooz and M. Rozali, ``String Dualities from Matrix
Theory,'' hep-th/9705175.}%
\refs{\rozali-\br}.
These theories also arise in the study  
of certain black holes in string theory 
\blackholes,
and it has been suggested that an improved understanding of some of
these theories may lead to progress in solving large $N$
nonsupersymmetric QCD \wittenfourone. 
The fixed points with 8 or
fewer supercharges are important in the problem of unifying M-theory
vacua, since they are crucial in connecting vacua with different
spectra of chiral fields
\nref\gh{O. Ganor and A. Hanany, ``Small $E_8$ Instantons and
Tensionless Noncritical Strings,'' hep-th/9602120, \np{474}{1996}{122}.}%
\nref\sw{N. Seiberg and E. Witten, ``Comments on String Dynamics in
Six Dimensions,'' hep-th/9603003, \np{471}{1996}{121}.}%
\nref\kschir{S. Kachru and E. Silverstein, ``Chirality-Changing
Phase Transitions in 4D String Vacua,'' hep-th/9704185.}%
\nref\aldazabal{G. Aldazabal, A. Font, L. E. Ibanez, A. M. Uranga and
G. Violero, ``Non-perturbative Heterotic $D=6$,$D=4$ $N=1$ Orbifold
Vacua,'' hep-th/9706158.}%
\refs{\gh - \aldazabal}.

Fixed point theories with $(2,0)$ supersymmetry
in six dimensions \refs{\wittentwoz,\strom} 
were recently studied in a matrix model formulation in \abkss.
The purpose of this paper is to move on to theories with $(1,0)$ 
supersymmetry
in six dimensions. We will formulate a matrix description of these
theories
and follow the chirality-changing phase transitions 
of \refs{\gh,\sw} 
in this language. 
We begin in section 2 with the definition of the
theory. In section 3 we analyze deformations away from the fixed
point, where we can see the low-energy spectrum in the spacetime
theory, and observe the chirality-changing phase transition. 
We discuss various interesting issues, which we are not able to fully 
resolve, concerning the matrix
description of these deformations.
Section 4
contains the $1+1$ dimensional generalization of the quantum
mechanical theory, which
corresponds to a six dimensional ``little string'' theory in spacetime.

As this paper was being completed, similar results
were independently obtained in \newlowe.

\newsec{The Quantum Mechanical Definition of the Fixed Point Theory}

We will study here the simplest example of a fixed point with
$(1,0)$ supersymmetry, which
is the low energy theory of a small instanton in the $E_8\times E_8$
heterotic string. In M theory this is described by a fivebrane 
at the end of the world ninebrane \refs{\gh,\sw}. 
This theory has a Coulomb branch
of the form $\IR/\IZ_2$ (times a decoupled $\IR^4$ factor),
on which the low energy spectrum consists
of a tensor multiplet and a hypermultiplet. The scalars in these
multiplets label the
transverse position of the fivebrane in M theory. The
scalar in the tensor multiplet parametrizes the distance between the
fivebrane and the ninebrane, and when its expectation value 
vanishes the low-energy theory is
superconformal. Another branch coming out of the superconformal point
is the Higgs branch, corresponding to enlarging the size of the
instanton. On this branch, the low-energy theory has 30 hypermultiplets,
which are in the $\half(\bf 56)+{\bf 1}+{\bf 1}$ representation
of the $E_7$ symmetry
left unbroken by the instanton. We would like to propose an
infinite momentum frame quantum mechanical description of this
theory, which  reproduces this moduli
space and low-energy spectrum. In particular,
we will consider in this framework 
the chirality-changing phase transitions
of \refs{\gh,\sw}. There is an obvious generalization of
this theory to $k$ coincident fivebranes (or small instantons), which
will also be discussed.

The arguments used in \abkss\ for the construction of $(2,0)$ theories
in six dimensions can also be used for the construction of theories
with $(1,0)$ supersymmetry.  To get a light-cone description of this
system, we start with M theory on $S^1/\IZ_2$ \hw\ with $k$
fivebranes, and compactify a longitudinal direction (of the
fivebranes) on a circle of radius $R$. The theory then becomes the
type IIA string theory on $S^1/\IZ_2$ (a.k.a. type $I^\prime$), 
with 8 D8-branes at each
orientifold fixed point \polwit\ and $k$ D4-branes.

In the next subsection we will
discuss the full matrix description of this theory. We will
introduce the degrees of freedom of the matrix description
of this system, their interactions, and their representations
under the various symmetries. In \S2.2
we will consider the limit
$M_p\to\infty$ in spacetime, and determine what remains of
the degrees of freedom in the matrix description in this limit.  
This surviving quantum mechanics is our formulation of the $(1,0)$ SCFT.

\subsec{Heterotic Fivebranes in Matrix Theory}

The above type $I^\prime$ system 
is equivalent to the $E_8\times E_8$ heterotic theory on a circle,
with a Wilson line $A_E$ breaking the gauge symmetry to $SO(16)\times SO(16)$
(and $k$ fivebranes wrapped around the circle).
Let the radius of this circle in the $E_8\times E_8$ theory be denoted $r_E$.
This vacuum is related by T-duality \ginsparg\ to the $SO(32)$ 
heterotic string 
on a circle of radius $r_S=1/4r_E$, with 
a Wilson line $A_S$ breaking the gauge group to $SO(16)\times SO(16)$.
The winding number $n_S$ of the $SO(32)$ theory maps to the 
D0-brane number
$N$ in the type I$^\prime$ description.  The $SO(17,1)$ T-duality
transformation maps this to a linear combination of
momentum, winding, and $E_8\times E_8$ lattice quantum numbers in the 
$E_8\times E_8$ theory:
\eqn\Tmap{N = n_{S}\leftrightarrow 2m_E-A_E^2n_E-2A_E\cdot P_E,}
where $m_E, n_E$ and $P_E$ are the momentum, winding, and $E_8\times E_8$
lattice quantum numbers in the $E_8\times E_8$ theory.

For the infinite momentum frame description
we are interested in states with large momentum $m_E$
around the circle in the $E_8\times E_8$ theory.  From
\Tmap\ we see that this corresponds to large D0-brane
number $N = n_S$, though the two quantum numbers are not
exactly the same.  

Let us now describe the quantum mechanics of the D0-branes in this
theory, near one of the orientifolds.
This quantum mechanics without the D4-branes was studied in
\nref\zeroeightorig{U. H. Danielsson and G. Ferretti, ``The Heterotic
Life of the D Particle,'' hep-th/9610082.}%
\nref\zeroeightks{S. Kachru and E. Silverstein, ``On Gauge Bosons in the 
Matrix Model Approach to M Theory,'' hep-th/9612162, \pl{396}{1997}{70}.}%
\nref\zeroeightlowe{
D.A. Lowe, ``Bound States of Type $I'$ D Particles and Enhanced Gauge
Symmetry,'' hep-th/9702006.}%
\refs{\zeroeightorig - \zeroeightlowe}. 
It is an $SO(N)$ gauge theory with 8 supersymmetries,
containing 16 fermions in the fundamental representation which arise
from the 0-8 strings. 
Adding the D4-branes (longitudinal fivebranes \berd) is done simply
by adding the 0-4 strings. These are
$k$ ``hypermultiplets'' in the fundamental representation,
and there is an $Sp(k) (\equiv USp(2k))$ 
global symmetry corresponding to these. For $N=1$
this theory was described in \douglas\ (see also \lambert). 
Altogether we are left
with four linearly realized supersymmetries, which is the 
correct number for a lightcone description of a spacetime theory with 8
supersymmetries. 

The global symmetry of the quantum mechanics is
\eqn\globsymm{SO(4)_\parallel \times SO(4)_\perp \times SO(16) \times
Sp(k),}
where $SO(4)_\perp$ corresponds to the rotation symmetry transverse
to the 4-branes (but inside the 8-branes), $SO(4)_\parallel$
corresponds to the rotations inside the 4-branes, $SO(16)$ is the
gauge symmetry on the 8-branes and $Sp(k)$ is the gauge symmetry
of the 4-branes. The 4 supersymmetry generators transform in the
$\bf \{(2,1) \ (2,1) \  1 \  1\}$ representation of this group, so that two
of its $SU(2)$ factors are in fact $R$-symmetries. The representations 
of the fields under the $SO(N)$ gauge symmetry and the global 
symmetries are given in the following table :
\eqn\reps{\matrix{ 
&&SO(N)&SO(4)_\parallel&SO(4)_\perp&SO(16)&Sp(k)\cr 
0-0\ {\rm states} : & A_0,X_9 & \bf N(N-1)/2 & \bf(1,1) & \bf(1,1)
& \bf{1} & \bf{1} \cr
& \alpha_L & \bf N(N-1)/2 & \bf(2,1) & \bf(2,1) & \bf{1} & \bf{1} \cr
& \beta_L & \bf N(N-1)/2 & \bf(1,2) & \bf(1,2) & \bf{1} & \bf{1} \cr
& X_\parallel & \bf N(N+1)/2 & \bf(2,2) & \bf(1,1) & \bf{1} & \bf{1} \cr
& \rho_R & \bf N(N+1)/2 & \bf(1,2) & \bf(2,1) & \bf{1} & \bf{1} \cr
& X_\perp & \bf N(N+1)/2 & \bf(1,1) & \bf(2,2) & \bf{1} & \bf{1} \cr
& \sigma_R & \bf N(N+1)/2 & \bf(2,1) & \bf(1,2) & \bf{1} & \bf{1} \cr
0-4\ {\rm states}: & v & \bf N & \bf(2,1) & \bf(1,1) & \bf{1} & \bf{2k} \cr
& \psi_R & \bf N & \bf(1,1) & \bf(2,1) & \bf{1} & \bf{2k} \cr
& \psi_L & \bf N & \bf(1,1) & \bf(1,2) & \bf{1} & \bf{2k} \cr
0-8\ {\rm states}: & \chi_L & \bf N & \bf(1,1) & \bf(1,1) & \bf{16} &
\bf{1}. \cr
}}
Here $X_{\parallel}$ gives the positions of the zero branes along
the fourbranes, $X_\perp$ gives the positions perpendicular to the
fourbranes, and $X_9$ gives the positions in the $S^1/\IZ_2$ direction.
In addition we have scalars $v$ in the fundamental representation.
The fermions (which are all real)
are denoted with subscripts R or L, according to their 
chirality in the corresponding $1+1$ dimensional theory of 1-branes, 
fivebranes and ninebranes, which is
related to the quantum mechanics we describe by
a T duality in the $x_9$ direction.
That theory has $(0,4)$ supersymmetry, and we will discuss it further
in section 4.
Supersymmetry pairs the right moving fermions with the bosons
appearing directly above them in the table, and $\alpha_L$ with the
gauge field. 

The moduli of the spacetime theory
are parameters in the quantum mechanics. These moduli 
are the scalars in the theory of the 4-branes and the 8-branes, which are
in the following 
representations :
\eqn\morereps{\matrix{
&&SO(N)&SO(4)_\parallel&SO(4)_\perp&SO(16)&Sp(k)\cr
4-4\ {\rm states}: & X^{(4)}_\perp & 
\bf{1} & \bf(1,1) & \bf(2,2) &
\bf{1} & \bf{2k(2k-1)/2} \cr
& X_9^{(4)} & \bf{1} & \bf(1,1) & \bf(1,1) & \bf{1} & \bf{2k(2k+1)/2} \cr
4-8\ {\rm states}: & H & \bf{1} & \bf(1,1) & \bf(2,1) & \bf{16} & \bf{2k} \cr
8-8\ {\rm states}: & X_9^{(8)} & \bf{1} & \bf(1,1) & \bf(1,1) & \bf{120} &
\bf{1}. \cr}}

Most of the interactions of this system may be easily derived from those
of the 0-brane/4-brane system, which is the dimensional reduction of a 6D 
$(1,0)$ theory,
and from those of the 0-brane/8-brane system 
\refs{\zeroeightorig - \zeroeightlowe}. 
Among the terms appearing in
the Lagrangian are terms 
of the following (schematic) form\foot{This
formula does not include the powers of the
gauge coupling $g_{QM}$, which may be put in on dimensional grounds.} :
\eqn\lagrangian{\eqalign{&\chi_L (X_9 - X_9^{(8)}) \chi_L + 
\psi_L (X_9 - X_9^{(4)})
\psi_L + \psi_R (X_9 - X_9^{(4)}) \psi_R + v^2 (X_9 - X_9^{(4)})^2 +
 \cr & \psi_L (X_\perp - X_\perp^{(4)}) \psi_R + 
v^2 (X_\perp - X_\perp^{(4)})^2 +
([X_\parallel,X_\parallel] + v^2)^2 +[X_\perp,X_\perp]^2+
v \sigma_R \psi_L + v \alpha_L \psi_R +\cr & 
\alpha_L [X_\parallel,\rho_R] + \beta_L [X_\parallel,\sigma_R] + 
\alpha_L [X_\perp,\sigma_R] + \beta_L [X_\perp,\rho_R] +
[X_\perp,X_\parallel]^2 + (H v)^2 + H \psi_R \chi_L. \cr}}
The singlet components of the fermions $\rho_R$ are completely 
decoupled, and their shifts
generate four non-linearly realized supersymmetries, completing the 8
spacetime supersymmetries. Quantization of the zero
modes of these fields will multiply the representation of each state
we get by $\{\bf(1,2)\ \bf(1,1)\} + \{\bf(1,1)\ \bf(2,1)\}$ (which is the
content of a half-hypermultiplet in spacetime).

In the quantum mechanics describing the superconformal point in
space-time, all the parameters \morereps\ vanish. Then,
the quantum mechanical theory has a Coulomb branch (in the usual sense
of a Born-Oppenheimer approximation) in which $X_\perp\ne 0$ and $v=0$.
In the matrix model interpretation, graviton states live here, as well
as $E_8$ gauge bosons localized near $X_9=0$
\refs{\zeroeightks,\zeroeightlowe}.
In addition, there is a Higgs branch in which $X_\perp=0$. It is
parametrized by expectation values of $X_\parallel$ and $v$, and
has (real) dimension 
\eqn\dimen{{\rm dim}~{\cal M}_H=
4Nk+4{{N(N+1)}\over 2}-4{{N(N-1)}\over 2}=4N(k+1).}

\subsec{Decoupling Gravity:  Formulation of the $(1,0)$ SCFT}

As in the case of the $(2,0)$ theories discussed in \abkss, the gauge
coupling in the quantum mechanics is related to the eleven dimensional
Planck mass $M_p$ and the compactification radius $R$ by $g_{QM}^2
\sim M_p^6 R^3$. Thus, taking $M_p \to \infty$ corresponds to the
$g_{QM} \to \infty$ (or IR) limit of the quantum mechanics, where we
expect the conformal theory in spacetime to decouple from gravity, as
well as from the $E_8$ gauge bosons whose gauge coupling goes to zero
in this limit. 

As in \abkss, the presence
of the Higgs branch (with no apparent spacetime interpretation)
is what signals the presence of the nontrivial conformal theory
in spacetime.  In the limit $g_{QM} \to \infty$, some degrees
of freedom become infinitely massive on the interior of
the Higgs branch.  In other words, the  
Coulomb and Higgs branches of the quantum mechanics decouple.
Integrating out the 0-4 states leads to an infinite tube on the
Coulomb branch, so the origin is infinitely far away on that branch
(where the gravitons and the $E_8$ gauge bosons live). 
The degrees of freedom from \reps\ that are lifted in the interior of 
the Higgs branch
($v\ne 0$, $X_\parallel\ne 0$) are :
\eqn\lifted{\eqalign{& A_0, X_9, \alpha_L, 2N(N-1) {\rm \ of\ the\
fields}\ 
v {\rm \ and\ } X_\parallel 
\ ({\rm and\ their\ superpartners}\ \psi_R, \rho_R), \cr
& X_\perp, \sigma_R, 2N(N+1) {\rm \ of\ the\ fields}\ \beta_L\ {\rm and}\ 
\psi_L.}} 
We are left with :
\eqn\surviving{\eqalign{&4N(k+1) {\rm \ of\ the\ fields\ } v {\rm \ and\ }
X_\parallel ({\rm and\ their\ superpartners}\ \psi_R,\rho_R), \cr 
&\chi_L, 4N(k-1) {\rm \
of\ the\ fields\ } \beta_L \ {\rm and}\ \psi_L. \cr}}
It is the $g_{QM}\to\infty$ theory of these degrees of
freedom that constitutes the matrix formulation
of the spacetime SCFT.
Note that, unlike in \abkss, even for $k=1$ there is a non-trivial
Higgs branch here. This corresponds to the non-trivial SCFT in
spacetime which exists even in this case.

The classical Higgs branch of the quantum mechanics
is the moduli space of $Sp(k)$ instantons \ndouglas. There is no
non-renormalization theorem for the moduli space in this case.  In the
quantum mechanics there could be loop corrections (say, involving the
$\chi_L$s) to the metric of this space. However, there is some fixed point 
governing the Higgs branch in the infrared ($g_{QM} \to \infty$) limit.  
We conjecture that, for $N \to \infty$, 
it correctly describes the $(1,0)$ superconformal
theories in the infinite momentum frame. 
In fact, it follows from the results of \finite\ that the
corresponding $1+1$ dimensional
$(0,4)$ sigma model, with target space ${\cal M}_{H}$ and with the 
additional left-moving fermion multiplets, is finite.
This is not to say that the infrared physics will necessarily
be transparent in terms of the degrees of freedom \surviving.
The IR theory may have complicated interactions, arising 
for instance, from the gauge constraint (the $A_0$ equation of motion)
in the original gauge theory we start from \zeroeightlowe. 

Note that since $N$ here is not the same as the spacetime momentum
$m_E$, the finite $N$ theory does not directly give us a discrete
light-cone description of the $(1,0)$ superconformal theories, as
suggested in \susskind. Presumably, as in \motls, the finite $N$
theory is a discrete light-cone quantization of these theories
compactified on a light-like circle with a Wilson loop breaking the
$E_8$ symmetry to $SO(16)$.
In the quantum mechanics only an
$SO(16)$ subgroup of the $E_8$ global symmetry of the $(1,0)$
superconformal theory in spacetime is visible.
As in \zeroeightks, the full $E_8$ representations get filled out as
the type I$^\prime$ coupling goes to infinity and
states of energy $1/\lambda_{I^\prime}$ come down.

\newsec{Low Energy States Away from the Fixed Point}

As discussed in the introduction, the six dimensional $(1,0)$ 
theories play a very
interesting role in giving chirality-changing phase transitions.
Within Lagrangian field theory there is no way to lift 
chiral matter, so it is interesting to consider how this
occurs in our formulation.
Let us perturb the spacetime theory away
from the conformal point, going into its Higgs or Coulomb branches.
Along these branches the low energy theory in spacetime is free, and
we should be able to find the correct low energy spectrum in our
quantum mechanical description. We will see how the quantum
numbers for these states arise in this section.

It is not clear to us that the deformed theory can be 
described using only the 
degrees of freedom \surviving\ that were involved in formulating
the critical theory.
In principle, there are two ways to analyze the theory away from the
conformal point. We could either perform the perturbation in the full
gauge theory and then take the $g_{QM} \to \infty$ limit (while
keeping the perturbation parameters finite), or work directly in the
theory which describes the Higgs branch of the quantum mechanics in
the $g_{QM} \to \infty$ limit, and analyze the perturbations in that
model.  As we will discuss in some detail below, we have
difficulty finding the correct spacetime spectrum using the
second approach.  

We interpret this difficulty as resulting from
the fact that this approach does not include information about
states localized at the singularities at the boundaries 
of the Higgs branch.
These quantum-mechanical
variables, though decoupled
from the interior of the Higgs branch at the conformal point, 
may still be important after we deform the theory away from the
conformal point.
Because of the tube metric, the
singularities of the Higgs branch still decouple from the
graviton/gauge boson
states which live on the Coulomb branch.  After turning on
the deformations \morereps, the quantum mechanical Higgs branch
is (generically) lifted, and the wave functions of all states 
are concentrated near the origin of the Higgs
branch. Thus, it is not a surprise that
the degrees of freedom 
related to the
singularities in the Higgs branch are required to describe the
states after the deformation.  
It would be interesting to understand better
the role of the singularities at the boundaries of the Higgs branch,
both in this theory and in the $(2,0)$ theories
described in \abkss.

\subsec{The Coulomb Branch}

First, let us discuss the Coulomb branch of the spacetime theory (this
is not to be confused with the Coulomb branch of the quantum mechanics).
On this branch the fivebranes are (generically) all separated from each
other and from the ninebrane.  There is a tensor multiplet and a
hypermultiplet (forming a tensor multiplet of $(2,0)$ supersymmetry) 
living on each fivebrane. For simplicity, let us focus
on the case $k=1$ (the other cases generically give $k$ copies of this). 
Moving into the Coulomb branch away from the critical
point is
done by turning on $X_9^{(4)}$, and we expect to find the fivebrane
states localized in the moduli space near $X_9 = X_9^{(4)}$
(specifically, when half of the eigenvalues of $X_9$ are equal to one of
the eigenvalues of $X_9^{(4)}$). In this
region the 0-8 strings are all massive and the $SO(N)$ gauge theory
is broken by the VEV of $X_9$ 
to $U(N/2)$ (here we take $N$ to be even). In the IR, the theory
reduces exactly to the quantum mechanics of D0-branes and D4-branes
(with 8 supersymmetries) discussed in \abkss, which is a
supersymmetric quantum mechanics on the moduli space of $U(k)$ 
instantons. In both cases the spacetime spectrum should include
a tensor multiplet and a hypermultiplet
for $k=1$ \foot{In the $(2,0)$ case it was not clear if a
$k=1$ theory which was decoupled from the Coulomb branch existed or
not \newwitten, but here we are reaching this theory by a perturbation
from a theory that was already decoupled from gravity, so there is no
problem.}. Thus, we should find 16 ground states of this theory,
which should be in the $\{\bf(1,3)\ \bf(1,1)\} + \{\bf(1,1)\ \bf(1,1)\} + 
\{\bf(1,1)\ \bf(2,2)\} + \{\bf(1,2)\ \bf(1,2)\} + \{\bf(1,2)\ \bf(2,1)\}$ 
representation of the $SO(4)_\parallel \times SO(4)_\perp$
global symmetry.  This representation arises by quantizing the fermion
zero modes of the $U(N/2)$-singlet components of the
fermions $\beta_L$ and $\rho_R$ appearing in table \reps. 

Note that in formulating the critical theory for $k=1$, we discarded 
$\beta_L$ because it became infinitely massive on the interior
of the Higgs branch \lifted.  But, as just noted, quantizing
its zero modes gives the correct degeneracy and quantum
numbers to describe the tensor multiplet on the spacetime
Coulomb branch.  This is the first difficulty we
find in attempting to describe the deformations away from
the critical theory using only the degrees of freedom involved
in formulating the fixed point itself.

In general, there is a correspondence between ground states of the
supersymmetric quantum mechanics on a space $X$ and cohomology
classes of $X$. Thus, we expect the modes of the tensor multiplet,
which should exist for any integer value of momentum around the
circle in the $E_8\times E_8$ theory
(i.e. for all even values of $N$ in the original $SO(N)$ theory),
to correspond to cohomology classes of the moduli space of our
theory. In the case $k=1$ and for non-zero $X_9$, this space is
simply the moduli space $\CM_{\tN}(U(1))$ 
of $\tN=N/2$ instantons in a $U(1)$ gauge group.
Since these instantons are necessarily all of zero size, this space
is just 
\eqn\modtriv{\CM_{\tN}(U(1)) = \IR^{4\tN} / S_{\tN}.} 

For $\tN=1$, we have simply a
0-brane/4-brane system, and the required state is just the bound state
of \dkps.   Indeed, this 
state becomes completely localized on the 4-brane in the
$M_p \to \infty$ limit.

For higher values of $\tN$, it is not apriori clear which cohomology
should be used, since the states are all associated with the
(orbifold) singularities of the moduli space.  It is natural
to conjecture that the
quantum mechanical ground states are given by the orbifold 
cohomology of this space \orb\ (this is more justified in the $1+1$
dimensional theory described in section 4, but our theory is just a
dimensional reduction of that theory). This gives a state for
every partition of $\tN$ \vafwit, in 
agreement with our expectation of finding
a single state for any integer value of momentum of the
tensor multiplet.  Quantizing 
the zero modes of $\beta_L$ and $\rho_R$ 
then gives this state the Lorentz quantum numbers of a
tensor multiplet and a hypermultiplet in the $(1,0)$
spacetime theory. These states are examples of states living at the
singularities of the Higgs branch, as discussed above.

\subsec{The Higgs Branch}

The other branch of the spacetime theory is the Higgs branch, in which
the fivebranes in spacetime turn into large $E_8$ instantons. 
We are only interested in the regime in which a quantum
field theory description, decoupled from gravity, remains
valid.  Let us denote by $\tilde H$ the canonically
normalized (dimension 2) scalar field in spacetime whose
VEV parameterizes the Higgs branch.  The field theory regime is
\eqn\regime{\tilde H<<M_{p}^2.}
On dimensional grounds, $\tilde H$ is related to the scale size $\rho$ 
of the instanton/fivebrane by
\eqn\scalesize{\tilde H=M_{p}^3\rho.}
Thus, the field theory regime is
\eqn\thin{\rho<<l_{p},}
where the fivebrane is still thin in Planck units.  
In the regime $\rho>l_{p}$, the fivebrane becomes thick,
gravity fails to decouple, and the matrix description 
necessarily involves the degrees of freedom \lifted\ as
well as \surviving.  In the field theory regime \thin,
as discussed above, one might hope to describe the theory
using only the degrees of freedom \surviving. However, as with
the spacetime Coulomb branch, we will encounter difficulties
in realizing this.

We will analyze
here 
only the case where 
the instantons are all embedded in a single $SU(2)$. 
In this case, the $E_8$ gauge symmetry in
spacetime is broken to $E_7$, and its $SO(16)$ subgroup (which appears
explicitly in the quantum mechanics) is broken to
$SO(12)\times SU(2)$. In the quantum mechanics, we go into this branch by
turning on the parameters corresponding to the 4-8 strings $H$ and to the
4-4 strings $X_\perp^{(4)}$. Note that turning on 
only the 4-8 strings when the
instantons are all in the same $SU(2)$ still leaves
all but one of the fivebranes/instantons at zero-size, 
so we still have a non-trivial
conformal theory for $k > 1$.  In the quantum mechanics we see that not
all of the Higgs branch is lifted in that case.  In contrast,
from \lagrangian\ we 
can easily see that turning on both $H$ and $X_\perp^{(4)}$ 
gives a mass to all the fields $v$ and $\psi_R$, and to $4k$ of the
fields $\psi_L$ and $\chi_L$. 
The first 12 components of $\chi_L$
(in the fundamental representation of the unbroken $SO(12)$ and of 
$SO(N)$) remain massless, as do 4 combinations of $\chi_L$ and
$\psi_L$ (again, in the fundamental of $SO(N)$). 

Naively, when we turn on $H$ the fields $v$ and $\psi_R$ become
massive, and there is no longer an infinite tube in the Coulomb branch
of the gauge theory, so gravity does not seem to decouple from our
theory. However, as discussed above, we should be careful in how we
normalize $H$. In spacetime, we want $\tilde H$ to remain finite as $M_p$
goes to infinity. 
This corresponds to having a finite $H$ in the
theory describing the Higgs branch in the $g_{QM} \to \infty$ limit.
In this limit, even for finite $H$
there is still an infinite tube in the Coulomb branch, 
and gravity still decouples from the
Higgs branch of the 6D SCFT.

For simplicity, we will analyze here only 
the case $k=1$, where the combinations that remain massless are exactly
the 4 fields $\psi_L$\foot{Since to get a free low-energy theory in 
spacetime for $k > 1$ we are
forced to turn on 4-4 strings, the general case decomposes in the IR
into $k$ copies
of this case (living at different values of $X_\perp$, corresponding to
the eigenvalues of $X_\perp^{(4)}$).}.
The hypermultiplet $H$ which obtains a VEV on the Higgs branch is
(using \morereps) charged under
$SU(2)_R\times SO(16)\times Sp(1)$, 
where $SU(2)_R$ is the first $SU(2)$ factor in $SO(4)_\perp$ (which
is identified with the $SU(2)_R$ symmetry of the spacetime
theory). Giving it a VEV breaks this symmetry to
$SU(2)_{R'}\times SO(12)\times SU(2)$, where $SU(2)_{R'}$ is
a subgroup of $SO(16)$ and $SU(2)_R$, and the last $SU(2)$ is a subgroup of
$SO(16)$ and $Sp(1)$ (but note that away from the small instanton
point this is a perturbative gauge symmetry from
the heterotic point of view). The fermions in the fundamental
representation of $SO(N)$ which remain massless are the $\chi_L$,
in the $\bf(1,12,1)$ representation, and $\psi_L$, in the $\bf(1,1,2)$
representation (and in the $\bf{2}$ of the other $SU(2)$ factor in
$SO(4)_\perp$).

Since the $v$ fields are all massive, the Higgs branch of the theory
after the perturbation is given simply by the space of $X_\parallel$s,
which is $\IR^{4N}/S_N$.
What states do we expect
to find in this case? The massless states of the spacetime theory on
the Higgs branch are 30 hypermultiplets. One of these hypermultiplets,
which corresponds to the transverse position of the instanton /
fivebrane (and
is free everywhere in the moduli space) is in the $\{\bf(1,1)\ 
\bf(2,2)\} + \{\bf(1,2)\ \bf(1,2)\}$ representation of the
$SO(4)_\parallel \times SO(4)_\perp$ global symmetry
(where $SO(4)_\perp$ now includes the new $SU(2)_{R'}$
group instead of the old $SU(2)_R$). The other hypermultiplets
are all in the
$2(\{\bf(1,1)\  \bf(2,1)\} + \{\bf(1,2)\ \bf(1,1)\})$ representation, and in
the $\half \bf{56} + \bf{1}$ representation of the unbroken $E_7$ gauge
group in spacetime. This representation decomposes into a 
$\half \bf(32,1) + \rm \half \bf(12,2) + \bf(1,1)$ of the 
$SO(12)\times SU(2)$ that we see in
the quantum mechanics.  According to \Tmap, the 
momentum modes of the first representation
should appear for odd values of $N$, while all the others should appear for
even values of $N$. Of course, this does not mean that the
momentum quantum number in the $E_8\times E_8$ string theory depends on the
representation: from \Tmap\ one sees that for $N=n_S$ odd,
$P_E$ must be a spinorial representation of $SO(16)\times SO(16)$
but $m_E$ can be odd or even.

As in the 0-8 system \refs{\zeroeightorig - \zeroeightlowe}, 
we expect the structure of the
ground states for odd (even)
values of $N$ to be the same as for $N=1$ ($N=2$), with the only change
being in the structure of the wave functions for the 0-8 bound states.
T duality and S duality relate our system to a heterotic $SO(32)$
string theory with some non-trivial instanton bundle, and there we can
show that the appropriate states exist (the calculation is essentially
as in \dg, and the presence of torsion does not change the results in
this case \rohm).

Let us analyze first the case $N=1$. In this case, the moduli space
is just $\IR^4$, so we have only the ground state. The 12 remaining
fermions $\chi_L$
are completely free in this case
(since the gauge symmetry is just $O(1) \equiv \IZ_2$), so they have
zero modes.   On the other hand, as explained in \douglas, 
the $\psi_L$s are sections of the $SU(2)$ instanton bundle
that lives in the $X_\perp$ directions.  In this
background they do not contribute any additional degeneracy.
Quantization of the $\chi_L$ zero modes gives states in the $\bf{32}+
\bf{32'}$ representation of the $SO(12)$ group.  As in 
\refs{\zeroeightks,\zeroeightlowe},
imposing a $\IZ_2 \equiv O(1)$ gauge constraint removes half of these states
and leaves us just with a $\bf{32}$. Adding the $\rho_R$ zero modes turns
each of these states into a half-hypermultiplet in spacetime, so we get
exactly the expected spectrum of states for this value of $N$.

In fact, for $N=1$ we can find the right states also if we work only
with the degrees of freedom \surviving\ involved in the critical
theory.  Then $v$ and its superpartners, as well as four
of the fields $\chi_L$, are lifted by $H$, and quantizing the zero
modes of the remaining fermions $\chi_L$ and $\rho_R$ provides us with
the required ${1\over 2}\bf{32}$ hypermultiplets (after taking into
account the $\IZ_2$ constraint).

For $N=2$, the situation is more
complicated (as it was also in the 0-8 case), since the interactions
between the fields play an important role in constructing the states.
To realize these states in our formulation,
we turn the operators (including $\chi_L$ and
$\psi_L$) into
creation and annihilation operators (as in \zeroeightlowe). 
We expect the ground states in
the quantum mechanics to be the same as those in the corresponding
$1+1$ dimensional sigma model, where a level-matching constraint will
force us to have two $\chi_L$ or $\psi_L$ oscillators in the sector
where they are anti-periodic (and no states will arise from other
sectors).
In the quantum mechanics, there will be a gauge constraint
(analogous to the level-matching constraint of the heterotic string)
which will force the total charge of a state under the $SO(2)$ gauge
symmetry to be equal to one \zeroeightlowe.  
We expect to find ground
states of the form $\chi_L \psi_L \vert 0 \rangle$ (where
$\chi$ and $\psi$ are now creation operators),
multiplied by an appropriate wave function
which turns this state into an $SO(4)_\parallel \times
SO(4)_\perp$-singlet.  
These states will be in the $\bf(12,2)$
representation of the $SO(12)\times SU(2)$ global symmetry
corresponding to the remaining spacetime gauge symmetry, and again the
$\rho_R$ zero modes will turn them into half-hypermultiplets.
The 29th and 30th hypermultiplets will arise from states  
involving two $\psi_L$s (contracted to form a singlet of the $SU(2)$
gauge symmetry), again with an appropriate wave function for the rest
of the fields.
It would be interesting to perform the Born-Oppenheimer
calculations explicitly, and see that exactly states of this form
arise. 

The IR theory of the degrees of freedom \surviving\ is complicated in
this case, and we have not been able to find these states directly by
deforming that theory. Presumably, this is again a result of the
theory at the singularities of the Higgs branch mixing with the theory
describing the interior of the Higgs branch as we deform away from the
conformal point.

\newsec{String Theories for string Theories}

The Higgs branch of the quantum mechanics formulated above is expected
to describe the $(1,0)$ superconformal theory in spacetime. In \abkss,
a similar quantum mechanics described the $(2,0)$ superconformal
theories in spacetime.  The corresponding $1+1$ dimensional theory
(which gives the quantum mechanics upon dimensional reduction) was
conjectured \refs{\abkss,\newwitten} to correspond to the ``little
string'' theory of the type IIA NS fivebrane
\refs{\gstring,\moorelitt}, which reduces at low energies to the
superconformal theory. Similarly, we expect the $1+1$ dimensional
theory with $(0,4)$ supersymmetry to describe the ``little string''
theory of the heterotic $E_8\times E_8$ fivebrane, defined by the
limit $g_s \to 0$ in that theory \gstring.

The field content and interactions of this theory are the same as
those described above, with $X_9$ now becoming part of the $1+1$
dimensional gauge field. The only difference is that there are now 32
chiral fermions $\chi_L$, since we can no longer ignore the states of
the ``other wall'' (these states are also required for anomaly
cancellation). As in the Matrix theory descriptions of the heterotic
\nref\bm{T. Banks and L. Motl, ``Heterotic Strings from Matrices,''
hep-th/9703218.}%
\nref\lowehet{D.A. Lowe, ``Heterotic Matrix String Theory,''
hep-th/9704041, \pl{403}{1997}{243}.}%
\nref\rey{S.J. Rey, ``Heterotic M(atrix) Strings and their
Interactions,'' hep-th/9704158.}%
string \refs{\bm - \rey}, half of these fermions have periodic boundary
conditions and the other half have anti-periodic boundary
conditions. The $X_9$ positions of the D0-branes turn into the Wilson
loop around the circle, and half of the $\chi_L$ fermions
are massless when the value of this Wilson loop corresponds to the
D0-branes being at each of the two walls. However,
this theory should still have a parameter $X_9^{(4)}$,
corresponding to the $X_9$ position of the fivebranes\foot{As noted in
\newken, this parameter actually lives in the ``Coxeter block'' of
$Sp(k)$.}, and the $\psi$
fermions (as well as their bosonic partners) 
should only be massless when the Wilson loop is equal to the
eigenvalues of $X_9^{(4)}$. This is realized in the $1+1$ dimensional
field theory by having the boundary conditions for the $\psi$ fields
around the circle twisted by an arbitrary $X_9^{(4)}$ matrix (in the
adjoint representation of $Sp(k)$), namely 
\eqn\boundcond{\psi(x+2\pi) =
\exp(X_9^{(4)}) \psi(x)~.} 
The $v$s have similar boundary
conditions. Note that such boundary conditions are
not possible for the $\chi_L$ fields since a
potential would be generated if their boundary condition were
different \bss.

We conjecture that the Higgs branch of this theory, in the $g_{YM} \to
\infty$ and large $N$ limits, gives an infinite momentum frame
description of the
``little string'' theory of the heterotic $E_8\times E_8$ fivebrane at
zero coupling. At low energies this theory goes over to the quantum
mechanics of the previous sections, as expected. Note that
the spacetime theory in this case includes two strings even for a
single fivebrane, coming from the membranes stretched between the
fivebrane and the two end of the world ninebranes. The sum of the
tensions of these two strings is the heterotic string tension $M_s^2$,
but their ratio depends on the parameter $X_9^{(4)}$ described above.

\bigskip

\centerline{\bf Acknowledgments}\nobreak

We would like to thank T. Banks, 
J. de Boer, G. Moore and N. Seiberg for useful
discussions. O.A. and E.S. would like to thank the Aspen
Center for Physics for hospitality during the course of this
work. The work of O.A. is supported by DOE grant DE-FG02-96ER40559.
The work of M.B. is supported by NSF grant NSF PHY-9513835.
The work of S.K. is supported by NSF grant PHY-95-14797 
and a DOE Outstanding Junior Investigator Award.  
The work of E.S. is supported by the DOE under contract
DE-AC03-76SF00515.

\listrefs
\end